\documentclass[prb,groupedaddress,showpacs]{revtex4}
\usepackage[dvips]{graphicx}
\begin{document}
\title{Shallow donor wavefunctions and donor-pair exchange in silicon: {\it Ab
initio} theory and floating-phase Heitler-London approach}
\author{Belita Koiller, R.B.~Capaz}
\affiliation{Instituto de F{\'\i}sica, Universidade Federal do Rio de
Janeiro, 21945, Rio de Janeiro, Brazil}
\author{Xuedong Hu}
\affiliation{Department of Physics, University at Buffalo, the State
University of New York, Buffalo, NY 14260-1500}
\author{S. Das Sarma}
\affiliation{Condensed Matter Theory Center, Department of Physics,
University of Maryland, College Park, MD 20742-4111}
\date{\today}
\begin{abstract}
Electronic and nuclear spins of shallow donors in Silicon are attractive 
candidates for qubits in quantum computer proposals.  Shallow donor exchange
gates are frequently invoked to preform two-qubit operations in such
proposals.  We study shallow donor electron properties in Si within the
Kohn-Luttinger envelope function approach, incorporating the full Bloch
states of the six band edges of Si conduction band, obtained from {\it ab
initio} calculations within the density-functional and pseudopotential
frameworks.  Inter-valley interference between the conduction-band-edge
states of Si leads to oscillatory behavior in the charge distribution of
one-electron bound states and in the exchange coupling in two-electron states. 
The behavior in the donor electron charge distribution is strongly influenced 
by interference from the plane-wave and periodic parts of the Bloch functions.
For two donors, oscillations in the exchange coupling calculated within the
Heitler-London (HL) approach are due to the plane-wave parts of the Bloch
functions alone, which are pinned to the impurity sites.  The robustness of
this result is assessed by relaxing the phase pinning to the donor sites. 
We introduce a more general  theoretical scheme, the floating-phase HL, from
which the previously reported  donor exchange oscillatory behavior is
qualitatively and quantitatively confirmed.  The floating-phase formalism
provides a ``handle'' on how to theoretically anticipate the occurrence of
oscillatory behavior in electronic properties associated with electron bound
states in more general confining potentials, such as in quantum dots.
\end{abstract}
\maketitle

\section{Introduction}

Doping in semiconductors has significant technological impact.  As transistors
and integrated circuits decrease in size, the physical properties of the
devices are becoming sensitive to the actual configuration of
impurities.\cite{voyles}  A striking example is the proposal of donor-based
silicon quantum computer (QC) by Kane,\cite{Kane} in which the monovalent
$^{31}$P impurities in Si are the fundamental quantum bits (qubits).  This
intriguing proposal has created considerable recent interest in revisiting
the donor impurity problem in silicon, particularly in the Si:$^{31}$P
system.

Two-qubit operations for the donor-based Si QC architecture, which are
required for a universal QC, involve precise control over electron-electron
exchange\cite{herring64,HD} and electron-nucleus hyperfine interactions.  Such
control can presumably be achieved by fabrication of donor arrays with
accurate positioning and surface gates whose potential can be precisely
controlled.\cite{Obrien,encapsulation,implant,schenkel03}  However, we have
shown\cite{KHD1} that electron exchange in bulk silicon has spatial
oscillations on the atomic scale due to the valley interference arising from
the particular six-fold degeneracy of the bulk Si conduction band.  These
oscillations place heavy burdens on device fabrication and coherent control,
because of the very high accuracy requirement for placing each donor inside
the Si unit cell, and/or for controlling the external gate voltages.  

The potentially severe consequences of these problems for exchange-based Si QC
architecture motivated us and other researchers to perform further theoretical
studies, going beyond some of the simplifying approximations in the formalism
adopted in Ref.~\onlinecite{KHD1}, and incorporating perturbation effects due
to applied strain\cite{KHD2} or gate fields.\cite{wellard03}  Both these
studies, performed within the standard Heitler-London (HL)
formalism,\cite{slater} essentially reconfirm the originally reported
difficulties regarding the sensitivity of the electron exchange coupling to
donor positioning, indicating that these may not be completely overcome by
applying uniform strain or electric fields.  At this point it is clear that
the extreme sensitivity of the calculated exchange coupling to donor relative
position originates from interference between the plane-wave parts of the six
degenerate Bloch states associated with the Si conduction-band minima.
Theoretically, this effect is dictated by the HL description of the
two-electron singlet and triplet states, defined as properly symmetrized
combinations of single-particle ground-state functions, where the phases of
the Bloch states are pinned at each donor site.  

Our goal in the present study is to assess the robustness of the HL
approximation for the two-electron donor-pair states.  Specifically,  
we first examine the single donor properties in more detail by including the
complete Si conduction band Bloch functions.  The calculated single donor
electron charge density vividly illustrates the rapidly oscillatory (and
non-commensurate) nature of the donor electron properties.  We then
relax the phase pinning at donor sites and allow small phase shifts in the
plane-wave part of the Bloch functions, which could in principle 
moderate, even eliminate,
the oscillatory exchange behavior.  Within this more general theoretical
scheme, which we call the {\em floating-phase} HL approach, our main
conclusion is that, for all practical purposes, phase shifts are
energetically unfavorable for both singlet and triplet states.  The
previously adopted HL wavefunctions are thus found to be robust, and the
oscillatory behavior obtained in Refs.~\onlinecite{KHD1,KHD2,wellard03}
cannot be taken as an artifact.
  
The present paper is organized as follows: In the next section we review the
shallow donor problem in Si, fully incorporating the details of the Si band
structure.  We present {\em ab initio} results for the bulk Si 
conduction-band-edge charge densities associated with individual Bloch states
and single donor states.  In Sec.~\ref{sec:pair} we consider two
substitutional donors in bulk Si, and introduce the floating-phase HL
approximation.  The two-particle ground state energy is compared with that of
the standard (pinned-phase) HL states.  We present results for the donor
electron exchange between P donor pairs in Si in situations of practical
interest, consistent with the current degree of experimental control over
donor positioning\cite{Obrien,encapsulation,implant} for the fabrication of
Si QC.  
We also indicate how the floating-phase scheme may be useful for different
Si-based QC architectures.  Concluding remarks are presented in
Sec.~\ref{sec:conclude}. 
This work thus provides necessary theoretical support and pictures to
anticipated experimental studies on qubit exchange coupling in a Si matrix.

\section{Single Donor in Silicon}
\label{sec:single}

We determine the donor electron ground state using effective mass theory.  The
bound donor electron Hamiltonian for an impurity at site ${\bf R}_0$ is
written as
\begin{equation}
{\cal H}_0={\cal H}_{SV}+{\cal H}_{VO} \,.
\label{eq:h0}
\end{equation}
The first term, ${\cal H}_{SV}$, is the so-called single-valley Kohn-Luttinger 
Hamiltonian,\cite{Kohn} which includes the single particle kinetic energy, 
the Si periodic potential, and the screened impurity Coulomb potential 
\begin{equation}
V({\bf r})=-\frac{e^2}{\epsilon|{\bf r}-{\bf R}_0|} \,. 
\label{eq:coul}
\end{equation}
For shallow donors in Si, we use the static dielectric constant $\epsilon =
12.1$.  The second term of Eq.~(\ref{eq:h0}), ${\cal H}_{VO}$, includes the 
inter-valley coupling effects due to the presence of the impurity 
potential.\cite{baldereschi70}  

The electron eigenfunctions are written in terms of the six unperturbed Si
band edge Bloch states $\phi_\mu = u_\mu(\bf r) e^{i {\bf k}_{\mu}\cdot {\bf
r}}$:
\begin{equation}
\psi_{{\bf R}_0} ({\bf r}) = \sum_{\mu = 1}^6  \alpha_\mu F_{\mu}({\bf r}-
{\bf R}_0) \phi_\mu ({\bf r}, {\bf R}_0)
= \sum_{\mu = 1}^6  \alpha_\mu F_{\mu}({\bf r}-
{\bf R}_0) u_\mu(\bf r) e^{i {\bf k}_{\mu}\cdot ({\bf r}-{\bf R}_0)}\,.
\label{eq:sim}
\end{equation}
The phases of the plane-wave part of all band edge Bloch states are naturally
chosen to be pinned at ${\bf R}_0$, and the $\alpha_\mu$ expansion
coefficients, also called valley populations, are real.  
In this way the charge density at the donor site
[where the donor perturbation potential Eq.~(\ref{eq:coul}) is more
attractive] is maximum, thus minimizing the energy for $\psi_{{\bf R}_0}({\bf
r})$.
In Eq.~(\ref{eq:sim}), $F_{\mu}({\bf r}-{\bf R}_0)$ are envelope functions 
centered at ${\bf R}_0$, for which we adopt the anisotropic Kohn-Luttinger
form (e.g., for $\mu = z$, 
$F_{z}({\bf r}) = \exp\{-[(x^2+y^2)/a^2 + z^2/b^2]^{1/2}\}/\sqrt{\pi a^2 b}$).
The effective Bohr radii $a$ and $b$ are variational parameters chosen to
minimize $E_{SV} = \langle\psi_{{\bf R}_0}| {\cal H}_{SV} |\psi_{{\bf
R}_0}\rangle$, leading to $a=25$ \AA, $b=14$ \AA~ and $E_{SV} \sim -30$ meV
when recently measured effective mass values are used in the
minimization.\cite{KHD1}

The ${\cal H}_{SV}$ ground state is six-fold degenerate.  This degeneracy is
lifted by the valley-orbit interactions,\cite{baldereschi70,Pantelides} which
account for intervalley scattering effects and are included here in ${\cal
H}_{VO}$. 
Valley-orbit effects are conveniently represented by two types of intervalley
couplings ${{\cal H}_{VO}}_{\mu,\nu}$: For valleys at perpendicular
directions (e.g., $\mu=x$, $\nu=z$) we take the coupling ${{\cal
H}_{VO}}_{x,z} = -\Delta_C$ while for those in opposite directions (e.g.,
$\mu=z$, $\nu=-z$), ${{\cal H}_{VO}}_{z,-z}=-\Delta_C(1+\delta)$.  Of course
${{\cal H}_{VO}}_{\mu,\mu}=0$.  Taking $\Delta_C=2.16$ meV and $\delta=-0.3$
correctly reproduces\cite{KHD2} the ordering and relative splittings of the
lowest energy states manifold for P donors in Si: A ground state of $A_1$
symmetry, followed by a triplet of $T_1$ symmetry and by a doublet of $E$
symmetry. In unstrained Si, the nondegenerate $A_1$ ground state corresponds to all
$\alpha_\mu = 1/\sqrt{6}$ in (\ref{eq:sim}), and its binding energy is
$E_0=<\psi_{{\bf R}_0}|{\cal H}_0|\psi_{{\bf R}_0}> = E_{SV}-(5+\delta)
\Delta_C \sim -40$ meV, to be compared to the experimental value\cite{madelung} of $-45$
meV.\cite{foot1}  
Aiming at the ground state of the system, we restrict our discussion to the 
nondegenerate ($A_1$-symmetry) ground state, thus all $\alpha_\mu =
1/\sqrt{6}$ in (\ref{eq:sim}) in relaxed Si.
For strained Si, also considered below, the valley populations change according to the
degree of strain.\cite{KHD2} 

The periodic part of each Bloch function is pinned to the lattice, independent
of the donor site.  It can be expanded over the reciprocal lattice vectors
${\bf G}$:
\begin{eqnarray}
u_\mu({\bf r})= \sum_{\bf G} c^\mu_{\bf G} e^{i{\bf G} \cdot {\bf r}}.
\label{eq:CK}
\end{eqnarray}
We determine the coefficients $c^\mu_{\bf G}$ for the conduction band edge
Bloch states in Eq.(\ref{eq:CK}) from {\it ab initio} calculations. 
Electron-electron interactions are described by density-functional
theory (DFT) within the local-density approximation
(LDA). \cite{hohenberg,kohn2} We use the exchange-correlation potential parametrized by
Perdew and Zunger \cite{perdew} from Ceperley-Alder quantum Monte-Carlo
results for the homogeneous electron gas \cite{ceperley}. The interactions
between valence electrons and ions are described by the {\it ab-initio},
norm-conserving pseudopotentials of Troullier-Martins \cite{tm}, 
generated by the FHI98PP code.\cite{fhipp} We use 290 plane waves in
the expansion of Eq.(\ref{eq:CK}), up to a maximum kinetic energy
of 16 Ry. Calculations are performed by the ABINIT code.\cite{abinit}
The key ingredients of this code are: (i) An efficient Fast Fourier Transform 
algorithm\cite{fft} for the conversion of wavefunctions between real 
and reciprocal space; (ii) The use of iterative minimization techniques to solve
the Kohn-Sham eigenvalue problem, more specifically an adaptation to a fixed 
potential of the band-by-band conjugate gradient method,\cite{cg} and a potential-based 
conjugate-gradient algorithm for the determination of the
self-consistent potential.\cite{pbcg} Details of the methodology are described
in Ref.~\onlinecite{cg}. We find the equilibrium lattice 
constant of Si at a$=5.41$ \AA~ and the conduction-band minima 
at $0.844 (2\pi/{\rm a})$ from $\Gamma$, in close
agreement with experimental results.\cite{madelung}  These values are used in
the calculations presented below.

In Figs.~\ref{fig:bloch}(a) and ~\ref{fig:bloch}(b) we present the electronic
probability density $\rho_x=|\phi_x|^2$ obtained from the {\em single}
conduction-band-edge Bloch state $\phi_x = u_x({\bf r}) e^{i{\bf k}_x\cdot{\bf
r}}$.  Visually, our results indicate that this state is predominantly formed
by $|p_x\rangle$ atomic-like orbitals, although some $d-$character may
also be present, in consistency with the higher degree of delocalization for
these states as compared to the conduction band states at the $\Gamma$
point.\cite{Richardson}  Of course, as for any Bloch state, the probability
density is periodic in the fcc lattice.  
It is also interesting to note that, among the 290 plane-wave states included in our 
basis, over 90\% of the spectral weight in the plane-wave expansion of $u_x({\bf
r})$ in Eq.(\ref{eq:CK}) comes from five reciprocal lattice vectors: ${\bf G}
= (0,0,0),~{\frac{2\pi}{\rm a}}(-1,\pm 1,\pm 1)$.  These give the five
smallest values of $|{\bf G}+{\bf k}_x|$ since $k_x = 0.844 \frac{2\pi}{\rm
a}(+1,0,0)$.  This same criterion for the five most relevant coefficients
$c^{\mu}_{\bf G}$ applies to each of the other five ${\bf k}_{\mu}$-vectors.\cite{foot2}

In Figs.~\ref{fig:bloch}(c) and \ref{fig:bloch}(d) we show the total charge
density $\sum_{\mu=1,6}|\phi_{\mu}|^2$.  Fig.~\ref{fig:bloch}(c) shows the
characteristic antibonding signature of the conduction band state, which was
also found by Richardson and Cohen \cite{Richardson} for the conduction-band
density at the $X$-point in Si (thus not exactly at the band edge).
The conduction band edge state of Si has been previously studied by Ivey and 
Mieher.\cite{ivey75}  Our {\em ab initio} results are in good qualitative 
agreement with this earlier empirical pseudopotential study.
\begin{figure}
\includegraphics[width=4.1in]
{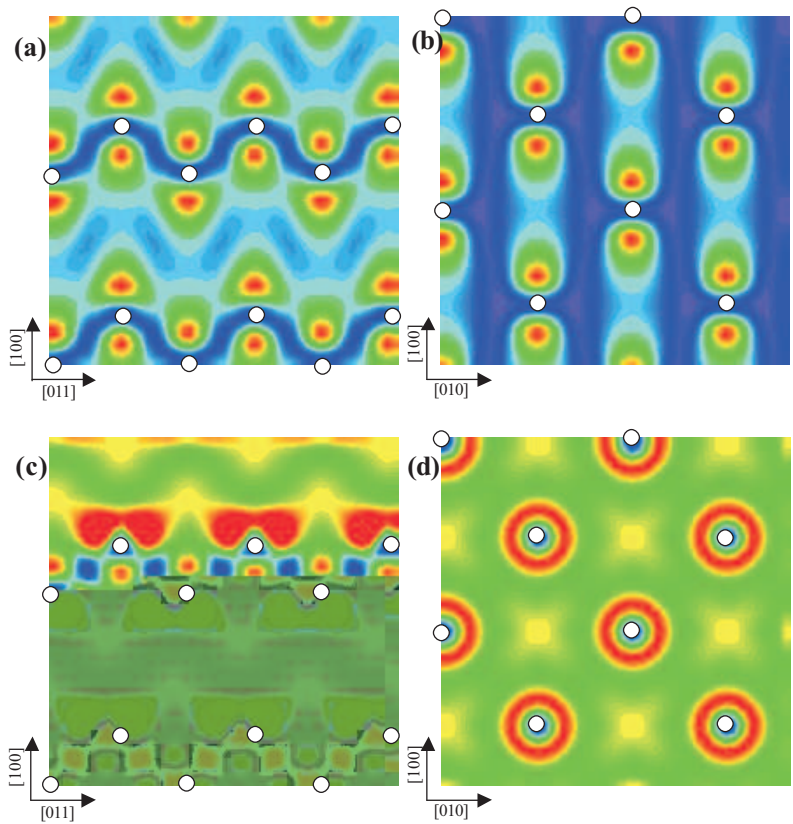}
\protect\caption[Four frames with charge distribution]
{\sloppy{
(Color) (a) and (b) Probability density for the single Bloch state 
$\rho_x=|\phi_x|^2$ in two different crystal planes. Notice the $|p_x>$ 
atomic-like signature. (c) and (d) Total probability density for the six 
conduction-band minima, showing a more symmetric structure. 
White dots represent Si sites in the diamond structure and the color 
scheme runs from purple (low density) to red (high density)}}
\label{fig:bloch}
\end{figure}

We analyze the effects of Si band structure on donor electron wavefunction 
within two models for the conduction band edge states $\{\phi_\mu\}$ of
Si: (i) $\phi_{\mu,{\bf R_0}} = e^{i {\bf k}_{\mu}\cdot ({\bf r}-{\bf R_0})}$; 
(ii) $\phi_{\mu,{\bf R_0}} = u_\mu({\bf r}) e^{i {\bf k}_{\mu}\cdot({\bf
r}-{\bf R_0})}$.  Model (i) corresponds to the free-electron
single-plane-wave-per-valley approximation adopted in previous
studies.\cite{KHD1,KHD2,Andres}  In model (ii) band structure contributions
are fully incorporated.  Regarding the electron probability density plotted
in Fig.~\ref{fig:bloch} for model (ii), model (i) would have given completely
uniform distributions, consistent with taking $u_\mu = 1$, i.e., $c^{\mu}_{\bf
G}=\delta_{{\bf G},\Gamma}$ for all $\mu$.  

\begin{figure}
\includegraphics[width=4.1in]
{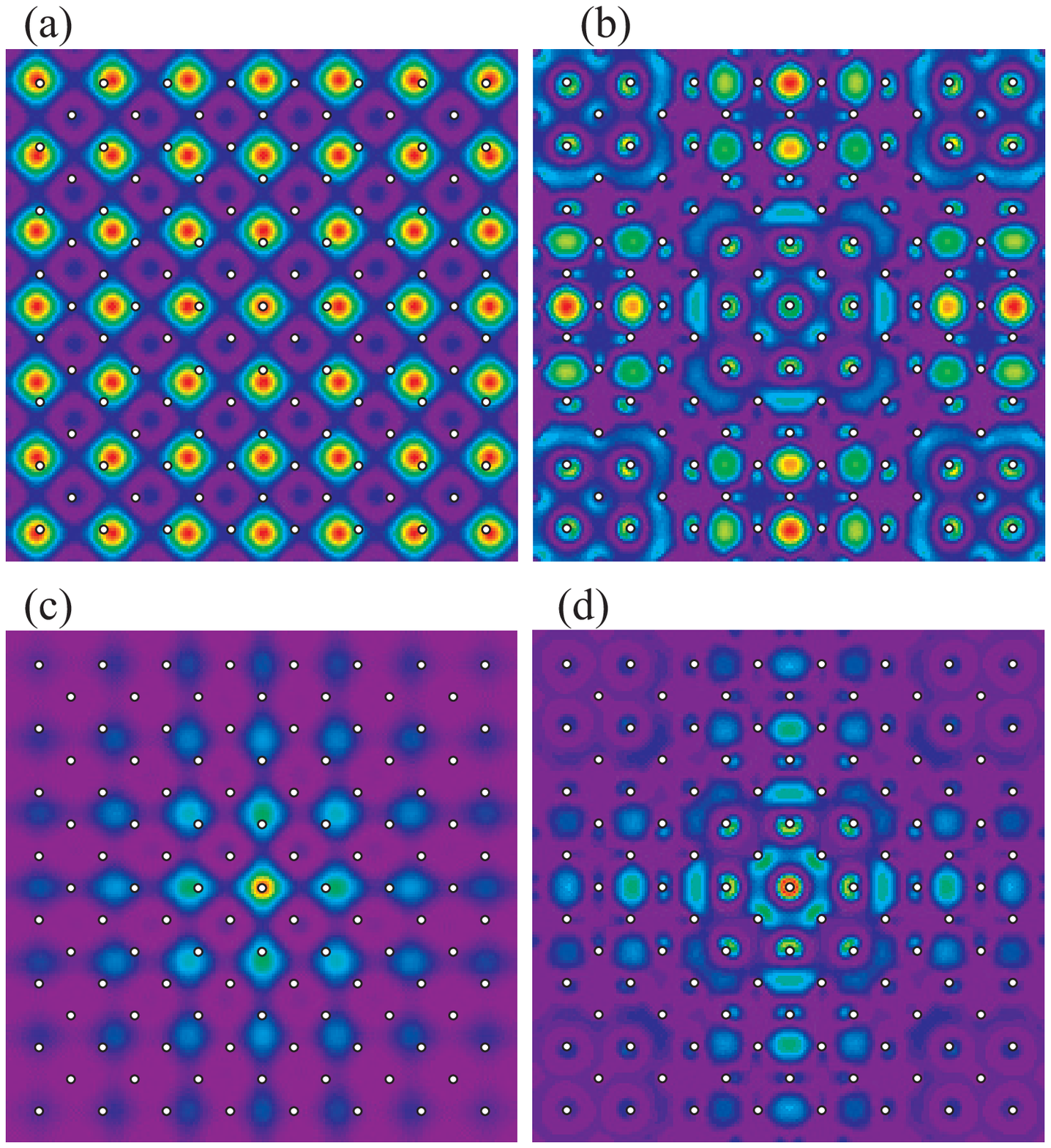}
\protect\caption[Four frames with charge distribution]
{\sloppy{
(Color) Frames (a) and (b) give the electron probability densities on the
(001) plane of bulk Si for the bottom of the conduction band eigenstate
corresponding to a symmetric combination of the six degenerate Bloch states
at the conduction band edge, calculated within models (i) and (ii)
respectively.  Frames (c) and (d) give the corresponding probabilities for
the ground state of a donor in Si within the envelope function
approximation.  The white dots give the in-plane atomic sites
and the color scheme runs from purple (low density) to red (high density)
}}
\label{fig:bands}
\end{figure}

The effects of the conduction band states of Si on the donor wavefunctions and
charge density are well established experimentally.\cite{Feher} 
Particularly, the single impurity charge density is not only an interesting
physical property by itself, but also foretells the oscillatory behavior in
two-donor properties such as exchange.  Figs.~\ref{fig:bands} (a) and (b)
give the single electron charge density $|\Phi ({\bf r})|^2$ along a (001)
crystal plane for a symmetrized 
state at the conduction-band edge of {\it bulk silicon},
$\Phi({\bf r}) = (\sqrt{6})^{-1} \sum_{\mu = 1}^6 \phi_{\mu,{\bf R}_0}(\bf
r)$, within models (i) and (ii) respectively.  Frame (a) shows that
interference from the six plane-wave states included in model (i) leads to a
periodic charge pattern consistent with a simple-cubic lattice of lattice
parameter $2\pi/k_{\mu}\sim 1.18$a, with a periodicity which is clearly
different (and incommensurate) from the atomic positions in the lattice,
since $|{\bf k}_\mu|$ is incommensurate with the reciprocal lattice.  Of
course a different interference pattern would result if the plane waves were
not {\em all} pinned at site ${\bf R}_0$.  Results for model (ii) given in
(b) show that additional interference from the Bloch functions $u_\mu({\bf
r})$, which are periodic in the fcc lattice, further reduce the periodicity
of the charge density.

Figures~\ref{fig:bands}(c) and \ref{fig:bands}(d) give the charge density
$|\psi_{{\bf R}_0} ({\bf r})|^2$ for the donor state in Eq.~(\ref{eq:sim}), 
within models (i) and (ii) respectively.  The impurity site ${\bf R}_0$,
corresponding to the higher charge density, is at the center of each frame.
It is interesting (and somewhat counterintuitive) that, except for this
central site, 
regions of high charge concentration and atomic sites do not necessarily
coincide,
because the charge distribution periodicity
imposed by the plane-wave part of the Bloch functions is $2\pi/k_{\mu}$,
incommensurate with the lattice period a.

\section{Donor pair}
\label{sec:pair}

\subsection{Exchange coupling within the Heitler-London approach}

Within the HL approximation, the lowest energy singlet and triplet
wavefunctions for two electrons bound to a donor pair at sites $\mathbf{R}_A$
and $\mathbf{R}_B$ are written as properly symmetrized combinations of
$\psi_{\mathbf{R}_A}$ and $\psi_{\mathbf{R}_B}$, which are in turn defined in
Eq.~(\ref{eq:sim})
\begin{equation}
\Psi^s_t({\mathbf r}_1,{\mathbf r}_2) = \frac{1}{\sqrt{2(1\pm S^2)}}
\left[ \psi_{{\bf R}_A}({\bf r}_1) \psi_{{\bf R}_B}({\bf r}_2) \pm 
\psi_{{\bf R}_B}({\bf r}_1) \psi_{{\bf R}_A}({\bf r}_2) \right],
\label{eq:hl}
\end{equation}
where $S$ is the overlap integral and the upper (lower) sign corresponds to
the singlet (triplet) state.  The energy expectation values for these states
are
\begin{equation}
E^s_t = \langle\Psi^s_t|{\cal H}|\Psi^s_t\rangle = 2E_0 + \frac{H_0 \pm H_1}{1
\pm S^2}
\label{eq:energies}
\end{equation}
where $E_0$ is the isolated impurity binding energy and $H_0$ and $H_1$ are
usually referred to as Coulomb and exchange integrals.\cite{KHD2,slater}
The energy difference $J=E_t-E_s$ gives the exchange splitting.
We have previously derived the expression for the donor electron
exchange splitting in Ref.\onlinecite{KHD2}, which we reproduce here:
\begin{eqnarray} 
J({\bf R}) = \sum_{\mu, \nu} |\alpha_\mu|^2 |\alpha_\nu|^2 {\cal J}_{\mu \nu}
({\bf R}) \cos ({\bf k}_{\mu}-{\bf k}_{\nu})\cdot {\bf R}\,,
\label{eq:exch}
\end{eqnarray}
%
where $\alpha_\mu$ are the valley populations defined in Eq.~(\ref{eq:sim}), and 
${\cal J}_{\mu \nu} ({\bf R})$ are kernels determined by the envelopes.
These are slowly varying functions of ${\bf R}$ (explicit expressions 
are given in Ref.~\onlinecite{KHD2}),  
monotonically decaying with distance since only the exponential envelopes centered 
at each donor, but not the Bloch functions, contribute to it. 
Below we make a few observations before we attempt to go beyond the HL approximation.

Equation~(\ref{eq:exch}) does not involve any contribution from the periodic
part of the Bloch functions (\ref{eq:CK}) 
[in terms of additional oscillatory behavior in $J({\bf R})$ or additional
contribution to the magnitude of $J({\bf R})$],
which therefore may essentially be taken as $u_{\mu}({\bf r})=1$.  This fact
has been pointed out by Wellard et al,\cite{wellard03} and is a consequence
of the pinning of the $u_{\mu}({\bf r})$ functions to the lattice,
independent of the donor site, and of their fast oscillating nature.  These
authors calculated some HL integrals with ${\bf G}$ different from the
$\Gamma$ point, which were originally neglected in
Refs.~\onlinecite{KHD1,KHD2,Andres}, and confirmed numerically that all
approximations adopted here (and in Ref.~\onlinecite{KHD2}) are excellent. 
We therefore conclude that models (i)
and (ii), though giving quite different electron probability densities as
illustrated in panels (c) and (d) of Fig.~\ref{fig:bands}, effectively lead to
the same results for the exchange coupling within the HL approximation.  

Although the exchange coupling given in Eq.~(\ref{eq:exch}) should be 
applicable to any relative position vector $\bf R$, including the effect of
small perturbations in the donor sites into off-lattice positions,\cite{KHD1}
it has been pointed out by Altarelli and
co-workers\cite{altarelli79,weiler84} that interstitial donors in Si may
acquire a deep-center character, invalidating the envelope-function treatment
adopted here.  We therefore focus our study on substitutional (thus shallow) 
donors in Si.  

\begin{figure}
\includegraphics[width=4in]
{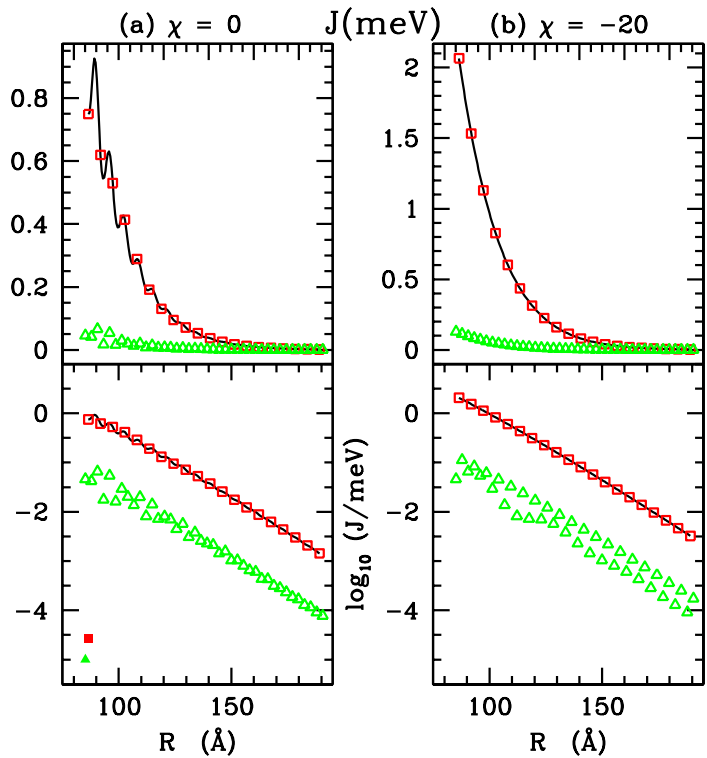}
\protect\caption[exchange with nn shifts]
{\sloppy{
(Color online) Calculated exchange coupling for a donor pair  versus interdonor distance in 
(a) unstrained and (b) uniaxially strained (along $z$) Si.
The open squares correspond to substitutional donors placed exactly along the 
[100] axis,  the lines give the calculated values for continuously 
varied interdonor distance along this axis, assuming the envelopes 
do not change. The open triangles give the exchange for a substitutional 
pair {\it almost} along [100], but with one of the donors displaced 
by $d_{NN}\sim 2.3$~\AA~ into a nearest-neighbor site. 
The lower frames give the same data in a logarithmic scale.
When the floating-phase HL approach is adopted, the results change negligibly; 
the filled symbols on the lower left frame 
give examples of calculated corrections (see text).
}}
\label{fig:nearest}
\end{figure}

Figure~\ref{fig:nearest} illustrates a case of practical concern involving
unintentional donor displacements into nearest-neighbors sites, when donor
pairs belong to different fcc sublattices.\cite{foot}  The open squares in
Fig.~\ref{fig:nearest}(a) give $J({\bf R})$ for substitutional donors along
the [100] axis, while the open triangles illustrate the different-sublattice
positioning situation, namely ${\bf R} = {\bf R_0} + {\vec \delta}_{NN}$ with
${\bf R_0}$ along the [100] axis and ${\vec \delta}_{NN}$ ranging over the 
four nearest-neighbors of each ${\bf R_0}$
($d_{NN}=|\vec\delta_{NN}|=$a$\sqrt{3}/4\sim 2.34$~\AA).  The lower panel 
of the figure presents the same data on a logarithmic scale, showing that
nearest-neighbor displacements lead to an exchange coupling reduction by one
order of magnitude when compared to $J({\bf R_0})$. 

Our previous studies\cite{KHD2} show that the extreme sensitivity of $J({\bf
R})$ to interdonor positioning is eliminated for on-lattice substitutional
impurities in uniaxially strained Si (e.g. along the $z$ axis) commensurately
grown over Si$_{1-x}$Ge$_x$ alloys {\it if {$\bf R$} remains parallel to the
interface $x$-$y$ plane}.  The strain is accommodated in the Si layer by
increasing the bond-length components parallel to the interface and 
decreasing those along $z$, breaking the cubic symmetry of the lattice and
lowering the six-fold degeneracy of the conduction band minimum to two-fold.
In this case, the valley populations $\alpha_\mu$ in the donor electron
ground state wave function (\ref{eq:sim}) are determined from a scalar valley
strain parameter $\chi$, which quantifies the amount of strain.  
Figure~\ref{fig:nearest}(b) gives $J({\bf R})$ in uniaxially strained (along
$z$ direction) Si for $\chi=-20$ for the same relative positioning of the
donor pairs as in Fig.~\ref{fig:nearest}(a).  Notice that the exchange
coupling is enhanced by about a factor of 2 with respect to the relaxed Si
host, but the order-of-magnitude reduction in $J$ caused by displacements of
amplitude $d_{NN}$ into nearest-neighbor sites still persists as ${\vec
\delta}_{NN}$ is not parallel to the $x$-$y$ plane.

\subsection{Floating-phase Heitler-London approach}

\subsubsection{Formalism}
In Refs.~\onlinecite{KHD1} and \onlinecite{KHD2}, as in the standard HL
formalism presented above, it is implicitly assumed that the phases $e^{-i
{\bf k}_{\mu}\cdot {\bf R}_0}$ in Eq.~(\ref{eq:sim}) remain pinned to the
respective donor sites $\mathbf{R}_0 = \mathbf{R}_A$ and $\mathbf{R}_B$, 
as we adopt single donor wavefunctions to build the two-electron wavefunction.
Although phase pinning to the donor substitutional site is required for the
ground state of an isolated donor ($A_1$ symmetry) in order to minimize
single electron energy, this is not the case for the lower-symmetry problem
of the donor pair.  In order to minimize the energy of the two-donor system,
here we allow the phases to shift by an amount $\mathbf{\delta R}$ along the
direction of the interdonor vector $\mathbf{R} = \mathbf{R}_B-\mathbf{R}_A$,
so that the single-particle wavefunctions in Eq.~(\ref{eq:hl}) become
\begin{equation}
\psi_{{\bf R}_A} ({\bf r}) = \frac{1}{\sqrt{6}} \sum_{\mu = 1}^6 F_{\mu}({\bf
r} - {\bf R}_A) u_\mu(\bf r) e^{i {\bf k}_{\mu}\cdot ({\bf r}-{\bf R}_A +
{\mathbf \delta R})}
\label{eq:sima}
\end{equation}
and
\begin{equation}
\psi_{{\bf R}_B} ({\bf r}) = \frac{1}{\sqrt{6}} \sum_{\mu = 1}^6 F_{\mu}({\bf
r} - {\bf R}_B) u_\mu(\bf r) e^{i {\bf k}_{\mu}\cdot ({\bf r}-{\bf R}_B -
{\mathbf \delta R})}\,.
\label{eq:simb}
\end{equation}
All terms appearing in Eq.~(\ref{eq:energies}) are now functions of
$\mathbf{\delta R}$, which we take as a variational parameter here, chosen
independently as $\mathbf{\delta R}_s$ and $\mathbf{\delta R}_t$ to minimize
$E_s$ and $E_t$ (since singlet and triplet states are orthogonalized through
the spin part of the wavefunction).  A similar {\em ansatz}, the so-called
floating functions approach, was suggested by Hurley\cite{hurley} as an
improvement over HL for the H$_2$ molecule, with the atomic orbitals
symmetrically shifted towards each other.  When the amplitude of the shift is
taken as a variational parameter, energy reduction thus obtained leads to a
significantly better agreement with experiment for the hydrogen molecule
total energy.\cite{hurley}  Since the phases in Eq.(\ref{eq:sim}) are
responsible for the oscillatory behavior of the exchange coupling between
donor electrons in Si, this more general variational treatment might lead to
changes in the previously reported\cite{KHD1,KHD2,wellard03} behavior of the
two-donor exchange splitting $J=E_t-E_s$.

\subsubsection{One- and two-center contributions}

Adopting the floating-phase forms given in Eqs.~(\ref{eq:sima}) and
(\ref{eq:simb}) in the HL expression (\ref{eq:hl}) leads to a modified 
expression for the expectation value of the energy in Eq.~(\ref{eq:energies})
for the singlet and triplet states.
The term $2E_0$ on the right hand side of Eq.~(\ref{eq:energies}) gives the
single-particle single-center contributions from both (isolated) impurities,
which should be taken here as $E_A+E_B=2E_A$. For the present model
Hamiltonian, within the floating phase HL approximation, we get
\begin{eqnarray}
E_A(\mathbf{\delta R})
& = & \langle \psi_{{\bf R}_A}| {\cal H}_A |\psi_{{\bf R}_A} \rangle\\
& = & E_{SV}- \frac{\Delta_C}{3} \left[2 (\cos\phi_x + \cos\phi_y +
\cos\phi_z)^2 + \delta(\cos 2\phi_x + \cos 2\phi_y + \cos 2\phi_z) - 3
\right] \,, 
\nonumber
\label{eq:ea}
\end{eqnarray}
where $\phi_\mu=-\phi_{-\mu}={\bf k}_\mu \cdot \mathbf {\delta R}$.  The
parameters $\Delta_C$ and $\delta$ are defined in Sec.~\ref{sec:single}, where
their numerical values are also given.  As expected, for $\mathbf{\delta
R}=0$ the above expression leads to $E_A(0)=E_0= E_{SV}-(5+\delta)\Delta_C$, 
while $\mathbf{\delta R}\ne 0$ leads to
\begin{equation}
E_A(\mathbf{\delta R})=E_0+\frac{\Delta_0}{2}\ge E_0.
\label{eq:d0}
\end{equation}
The correction $\Delta_0$ is positive definite by construction, since the
one-particle functions in standard HL ($\mathbf{\delta R}=0$) are taken as the
ground-state wavefunction of the isolated impurity problem. 

The expectation value for the energy of the donor pair is given by
\begin{equation}
E_t^s(\mathbf{R},\mathbf{\delta R}) = E_t^s (\mathbf{R},0) + \Delta_0
(\mathbf{\delta R}) + \Delta_t^s (\mathbf{R},\mathbf{\delta R}) \,, 
\label{eq:all}
\end{equation}
where $E_t^s(\mathbf{R},0)$ is the pinned-phase result from the regular HL
calculation.  The first correction term, $\Delta_0 (\mathbf{\delta R})$, is
the energy shift due to the single-particle single-center contributions,
derived above, and $\Delta_t^s (\mathbf{R},\mathbf{\delta R})$ 
are the singlet and triplet state corrections coming from the two-center
contributions $H_0$, $H_1$ and $S$.  The latter are integrals involving the
electronic wavefunctions (\ref{eq:sima}) and (\ref{eq:simb}), and are
calculated here as described in Ref.~\onlinecite{KHD2}, with the proper phase
shifts included in the plane-wave part of the Bloch functions.

\begin{figure}
\includegraphics[width=4in]
{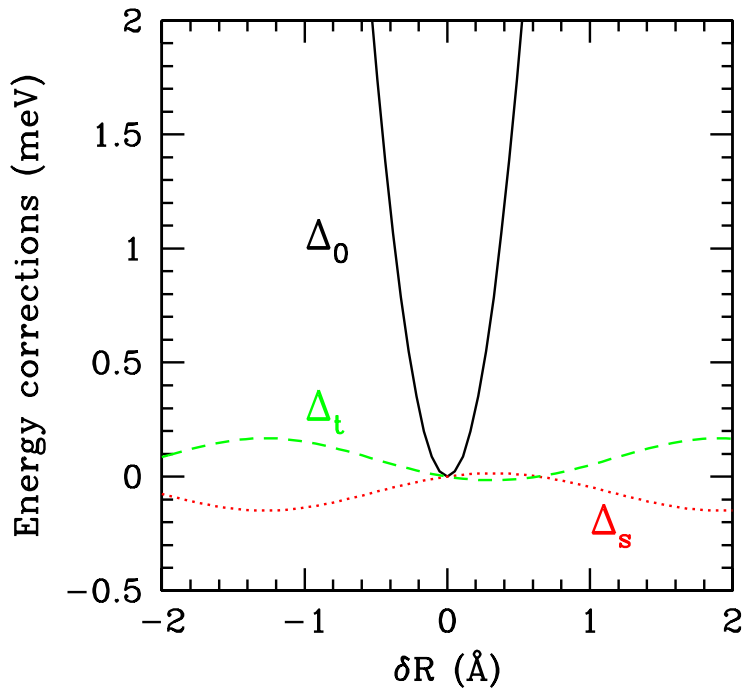}
\protect\caption[energy shifts versus delta R]
{\sloppy{
(Color online) Calculated corrections to the total energy for a P donor pair in Si. The
donors are 87~\AA~ apart, along the [100] direction.  The parameter $\delta
R$ is the amplitude of the individual phase shifts from the donor sites, $\pm
\mathbf{\delta R}$, along the interdonor line.  The solid line gives
$\Delta_0$, the single-center contribution, while the dotted (dashed) line
gives $\Delta_s$ ($\Delta_t$), the two-center singlet and triplet
contributions, respectively.
}}
\label{fig:delta}
\end{figure}

In Fig.~\ref{fig:delta} we give the calculated values of the individual 
corrections $\Delta_0 (\mathbf{\delta R})$ and $\Delta_t^s (\mathbf{R},
\mathbf{\delta R})$ for a geometry where the impurities are 16 lattice
constants apart $(\sim 87$ \AA), with $\mathbf R$ along the [100] crystal
direction.  The energy correction $\Delta_0$ raises sharply for nonzero
$\delta R$, and is of course independent of the relative position vector
${\bf R}$, while the energy variations $\Delta_t^s$ oscillate and decrease
with increasing relative distance $R$, and may be positive or negative
according to $\delta R$ (for $\delta R\approx 0$ in the case illustrated in 
Fig.~\ref{fig:delta}, $\Delta_s$ decreases for negative shifts $\delta R$,
while $\Delta_t$ decreases for positive shifts).  Since $\Delta_0$ is always
positive, independent of $R$ and very sensitive
to $\mathbf{\delta R}$, we conclude that the effect of phase shifts aiming at
minimizing two-donor energy is negligible and may be safely ignored for  
$R \gg a,b$, where $a$ and $b$ are the donor effective Bohr radii.  For
example, minimization of the total energy in Eq.~(\ref{eq:all}) for the
particular geometry considered in Fig.~\ref{fig:delta} leads to $\delta R_s =
-7$~m\AA, with the singlet energy decrease of 270~neV, and 
$\delta R_t = +7$~m\AA, with the triplet energy decrease of 6~neV.  
This results in an increase in $J$ by $(264)$~neV, given by the solid square
in the lower left hand side frame of Fig.~\ref{fig:nearest}.  The floating
phases variational scheme leads to a reduction in both singlet and triplet
states energy, therefore the net variation in $J$ is positive (negative) if
the triplet energy reduction is smaller (larger) than the singlet.  The solid
triangle in Fig.~\ref{fig:nearest} corresponds to a case of negative
variation, obtained when one of the donors in the above geometry is displaced
into a nearest-neighbor site.  Note that the corrections are more than three
orders of magnitude smaller than the calculated $J$ assuming $\delta R_s =
\delta R_t = 0$.  In other words, for all practical purposes the fixed-phase
standard HL approximation is entirely adequate for the range of interdonor
distances of interest for QC applications. 

This conclusion is not in contradiction with Hurley's result for the 
hydrogen molecule,\cite{hurley} where significant energy reduction is obtained
around the equilibrium nuclear separation, $R\sim 1.5 a_0$ ($a_0=0.53$ \AA~is
the free hydrogen atom Bohr radius).  For $R$ of the order of the Bohr
radius , $\Delta_s$ becomes comparable to $\Delta_0$, resulting in an
improved variational estimate for the ground state energy of the H$_2$
molecule when small shifts are allowed in the single-particle hydrogenic
orbitals.

Since the current calculation has taken into account the full bandstructure of
the host Si material, and modifying the standard HL approximation has proved
to have minimal effect on the results, further improvement in a perfect
crystal environment (that is, relaxed bulk Si) can only be achieved by
including the higher energy orbitals.\cite{HD,faulkner69}  However, we do not
anticipate significant moderation of the fast oscillatory behavior of
exchange coupling as all the orbitals share the same conduction band valleys,
though quantitative shifts might be expected in a larger scale molecular
orbital calculation.  In the present study, we keep the two donors relatively
far apart so that the HL approximation is applicable.  This is also the
situation of interest to practical QC fabrication considerations (which
requires the donors to be at least 100 \AA ~ apart).

Another improvement over our current calculation may come from including the
effect of lattice distortions.  The Coulomb interaction between the additional
protons on the lattice sites and the two electrons for the donor pair creates
a strain field on the underlying crystal lattice.  Such a field affects the
electronic structure in the same way as the uniaxial strain discussed above,
though it is along the inter-donor axis.  Since the inter-donor separation in
the present situation is much larger than the effective Bohr radius ($\sim
30$ \AA), the inter-donor interaction is strongly screened, therefore
lowering the strength of the strain. 
Furthermore, if a uniaxial strain is already applied along the $z$ direction,
so that the donor ground state only consists of the $z$ and $-z$
valleys,\cite{KHD2} the additional strain due to the presence of another
donor (e.g. along $x$ direction) will not further reduce the number of
valleys involved---the nondegenerate ground state will still consist of an
equally weighted superposition of these two valleys (instead of just one of
them), so that oscillations in exchange due to valley interference cannot be
removed.\cite{KHD2}
Nevertheless, a quantitatively analysis is needed to assess the significance
of this effect.  

\subsubsection{Coupled quantum dots}

Shallow donors in semiconductors may be viewed as the simplest, naturally
occurring quantum dot.  Compared to the gated quantum dots,
a relevant difference is the presence of a well defined and sharp pinning
center at the substitutional donor site.  Previous proposals of quantum dots
as quantum gates in a Si or Ge matrix\cite{levy01,friesen03} were based on
estimates for the exchange coupling within an {\it empty envelope function}
description.  It is clear that, for these materials, the plane-wave
parts of the Bloch functions may also have an important effect in the
exchange coupling. 

The floating-phase HL approach should be applicable to coupled quantum dots, 
leading to an expression equivalent to Eq.~(\ref{eq:all}).  The absence of a
sharp pinning center associated with each quantum dot implies that $\Delta_0$
is not as sensitive to the phase shifts in a floating-phase variational
scheme as obtained here for the donor case.  It is possible that variations
in the two-center contributions $\Delta^s_t$ dominate energetically and
determine the singlet and tripled ground state energies, whose difference
should give a reliable estimate for $J$.

Of course the valley-orbit effects described by ${\cal H}_{VO}$ in  
Sec.~\ref{sec:single}, which are quantitatively well established for P donors
in Si, would have to be estimated for the quantum dot confining potential,
including other perturbations which break the translational symmetry of the
host potential, such as the presence of nearby interfaces and 
strain.\cite{KHD2}  As in the present case, ${\cal H}_{VO}$ should lift the
six-fold degeneracy of the isolated quantum dot ground state.  An
investigation of valley-orbit effects in Si quantum wells was performed 
recently by Boykin {\it el al}.\cite{boykin} 

A similar scheme may also be useful for spin cluster qubits\cite{meier03} 
embedded in Si or Ge, where exchange gates are also invoked for inter-cluster
interactions. 
Demonstration that the exchange oscillatory behavior is circumvented for spin 
clusters would further require the formalism to be generalized to include
multielectron states\cite{HD1} in each cluster, as was explored in
Ref.~\onlinecite{Mizel}.

\section{Concluding remarks}
\label{sec:conclude}

We have included and assessed full band structure effects in the single donor
wavefunctions and charge distributions in Si.  We find interesting
oscillatory patterns resulting from interference between the different plane
wave components of the Bloch functions.  Regarding the well-separated donor
pair problem, we introduced a generalized scheme---the floating-phase HL
approach, which reconfirmed the reliability of standard HL for this range of
donor separations.

One perceived advantage\cite{Kane} of Si-based spin quantum computation 
(over, for example, the corresponding GaAs quantum dot based quantum
computation) is the universal nature of each qubit in Si, i.e. the fact that
the P donor electronic state in Si is always exactly the same, making each
qubit identical (without any need for additional characterization of
individual qubits which will surely be needed for GaAs quantum dot quantum
computers since electrostatically confined electronic spin states in GaAs
quantum dots would obviously have a fair amount of qubit to qubit variations
as no two quantum dots can really be identical).  Our finding of exchange
oscillations in Si donor states demonstrates that this perceived
advantage of Si comes with a price, where the exchange coupling between qubits
may vary depending on the precise positioning of the P atoms within the Si
unit cells.  We believe that, in spite of this problem, the QC scheme with
donors in Si still has its appeal in terms of uniform qubits.  
Obviously some characterization of the exchange coupling in Si becomes
necessary in view of the oscillatory exchange behavior.  We have discussed
elsewhere \cite{Raman03} how some precise local information about donor state
exchange coupling in Si can be obtained by using the powerful tool of the
micro-Raman scattering spectroscopy.  In addition, various band engineering
procedures,\cite{KHD2} using strain effects and/or Si-Ge quantum dots, could
be utilized to reduce the exchange oscillation effects, although its complete
elimination may not be easy.

From the perspective of current QC fabrication efforts, $\sim 1$ nm accuracy
in single P atom positioning has been recently
demonstrated,\cite{encapsulation} representing a major step towards the goal
of obtaining a regular donor array embedded in Si.
As expected, \cite{encapsulation} electronic calculations\cite{martins03} 
have confirmed that this degree of control is entirely compatible with the
operations involving the so-called A-gates in the Kane qubit
architecture.\cite{Kane}  On the other hand, the present calculations have
confirmed that deviations in the relative positioning of donor pairs with
respect to perfectly aligned substitutional sites along [100] lead to
order-of-magnitude changes in the exchange coupling.  Severe limitations in
controlling $J$ would come from ``hops'' into different fcc sublattices, in
particular among nearest-neighbor substitutional sites.  Therefore, precisely
controlling exchange gates in Si remains an open challenge.

\begin{acknowledgments}

We thank Alexei Kaminski for help with the preparation of the manuscript.
This work was partially supported by ARDA and LPS at the University of
Maryland, by ARDA and ARO at the University at Buffalo, and by Brazilian
agencies CNPq, FUJB, FAPERJ, PRONEX-MCT and Instituto do Mil\^enio de
Nanoci\^encias-CNPq. 
\end{acknowledgments}

\bibliography{float1}

\begin{thebibliography}{45}
\bibitem[{foo({\natexlab{a}})}]{foot1}
\bibinfo{note}{More sophisticated approaches such as multi-valley effective
  mass (MVEM) theories have been developed for impurities in
  Si.\cite{Pantelides} We do not adopt such an approach in order to maintain a
  simple and clear physical picture, as approaches like MVEM mostly provide
  better estimates for the binding energy, but do not seriously alter the
  valley interference features.}
\bibitem[{foo({\natexlab{b}})}]{foot2}
\bibinfo{note}{Our results presented below on donor electron exchange do not
  differ significantly from those in Ref.~\onlinecite{wellard03} (where 125
  plane-wave basis set was used), though we present in Figs.~\ref{fig:bloch}
  and \ref{fig:bands} more detailed information on the charge distributions.}
\bibitem[{foo({\natexlab{c}})}]{foot}
\bibinfo{note}{In MBE growth of donor arrays in Si, currently available
  technology does not allow differentiation for donor positioning among the two
  sites within one dimer (which are surface-nearest-neighbor to each other) of
  the Si (001) surface.\cite{Obrien} We believe a similar difficulty occurs for
  the 3D lattice nearest-neighbor sites.}
\end{thebibliography}

\end{document}